\documentclass[preprintnumbers,  11pt,  floatfix,
preprintnumbers,  letterpaper,  amsmath,  amssymb,  nofootinbib,
superscriptaddress]{revtex4}
\usepackage{graphicx}
\usepackage{epsfig}
\usepackage{bm}
\usepackage{amsfonts}

\begin{document}

\title{Non-singular Cyclic Cosmology without Phantom Menace}

\author{Yi-Fu Cai }
 \affiliation{ Department of Physics,   Arizona State
University,
Tempe,   AZ 85287,   USA \\ email: ycai21@asu.edu}

\author{Emmanuel N. Saridakis}
 \affiliation{ Center for Astrophysics, Space Physics and
Engineering Research,\\ Physics Department,
Baylor University,
Waco, TX  76798-7310, USA\\ email: Emmanuel$\_$Saridakis@baylor.edu}

\maketitle

 \subsection*{Abstract}
In this article we review recent developments of cyclic
cosmology. A typical non-singular cyclic model within General
Relativity requires a non-conventional fluid with negative
effective energy density,  in order to cancel the  matter
component and lead to a non-singular bounce. However,  the
existence of such a non-conventional fluid usually leads to
quantum instabilities and makes the theory ill-defined. In the
present work we follow the alternative way,  obtaining two
scenarios of non-singular cyclic cosmological evolutions in the
context of gravitational theories beyond General Relativity. The
degrees-of-freedom examination reveals that these two models
are free of the Phantom Menace. Our analysis illustrates that,  if
cyclic cosmology describes overall universe,  a theory of gravity beyond
Einstein may be expected.\\
\\
{\textbf{ Keywords:}} non-singular cosmology,  modified gravity,
bounce,  Horava-Lifshitz gravity

\newpage

\section{Introduction}

Big Bang cosmology,  considered as the ``standard model of the
universe",  has achieved numerous successes in explaining the overall
universe. However,  a crucial unexplainable issue of this paradigm is the
existence of the Big Bang itself,  which suggests that our
universe was born from a spacetime singularity. At that moment,
physical quantities like density, pressure, temperature, spacetime
curvature etc, were infinite and thus theoretically
ill-defined. Additionally, open geometries should be totally excluded,
since they correspond to infinite proper three-volume at the
singularity. As a consequence, there has been a lot of effort in
resolving this problem through quantum gravity effects or
effective field theory techniques.

A phenomenological solution to the cosmic singularity problem may
be provided by non-singular bouncing cosmologies
(Tolman 1931). Such scenarios have been constructed
through various approaches to modified gravity
(Mukhanov:1991zn),  such as the Pre-Big-Bang
(Veneziano 1991, Gasperini \& Veneziano 1993) and the
Ekpyrotic
(Khoury et al. 2001,  Khoury et al. 2002) models, gravity
with higher
order corrections (Brustein \& Madden 1998,  Biswas et al.
2006),  braneworld
scenarios (Shtanov \& Sahni 2003,  Saridakis 2009),
non-relativistic
gravity (Cai \& Saridakis 2009, Saridakis 2010), $f(T)$ modified gravity
(Chen et al. 2011, Cai et al. 2011b) and loop
quantum cosmology (Bojowald 2001). Non-singular bounces may be
alternatively investigated using effective field description in the
General Relativity background, introducing matter fields violating the null
energy condition, the so-called Phantom fields, leading to observable
predictions too (Cai et al. 2007, Cai et al. 2008),  introducing
non-conventional mixing terms (Saridakis \& Ward 2009,  Saridakis \&
Sushkov 2010, Saridakis \& Weller 2010), or in the frame of a close
universe (Martin \& Peter 2003). However, these General-Relativity
bounce-models suffer from the severe problem of quantum instability due to
the Phantom field
(Carroll et al. 2003, Cline et al. 2004), since once such a field with a
negative kinetic term is introduced the vacuum becomes unstable (note also
that the very concept of a negative kinetic energy itself sounds
unphysical). In summary, the extension to gravitational theories beyond
General Relativity seems to be the plausible and safest solution to the
singularity problem.

An interesting extension of bouncing scenarios is the (old)
paradigm of cyclic cosmology, in which the universe experiences
the periodic sequence of contractions and expansions
(Tolman 1934). Cyclic cosmology has gained new interest after the
appearance of Quasi Steady State cosmology  (Hoyle et al. 1993, Hoyle et
al. 1994), in which the C-field (Creation-field) is responsible for the
repeated cycles. Furthermore, it has been revived in the past
few years (Steinhardt \& Turok 2002a,  Steinhardt \& Turok
2002b),  since it
brings different insights for the origin of the observable
universe (see Novello \& Bergliaffa 2008 for a review).

Let us first analyze the general features of a cosmological
bounce and turnaround. The basic picture for the evolution of a
cyclic universe can be shown below:
\begin{eqnarray}
 ... {\rm bounce} \xrightarrow{expanding} {\rm turnaround}
\xrightarrow{contracting} {\rm bounce} ...~.\nonumber
\end{eqnarray}
Whether a universe is expanding or contracting depends on the
positivity of the Hubble parameter ($H\equiv\dot{a}/a$ with $a$
the scale factor of the universe). In the contracting phase that
exists prior to the bounce,  the Hubble parameter $H$ is
negative,
while in the expanding one that exists after it we have $H>0$. By
making use of the continuity equations it follows that at the
bounce point $H=0$. Finally,  it is easy to see that throughout
this transition $\dot H> 0$. On the other hand,  for the
transition from expansion to contraction,  that is for the
cosmological turnaround,  we have $H>0$ before and $H<0$ after,
while exactly at the turnaround point we have $H=0$. Throughout
this transition $\dot H < 0$.

An oscillating universe realized in spatially flat geometry
described by Einstein gravity has been studied in
(Xiong et al. 2008, Cai et al. 2010),  by making use of a
quintom
matter (Feng et al. 2005),  which involves a ghost degree of
freedom,  and thus it cancels the contribution of conventional
matter fields. Unfortunately,  these types of models suffer from
the problem of quantum instability due to an unbounded vacuum
state from below,  which is the so called ``Phantom Menace"
(Carroll et al. 2003, Cline et al. 2004).

Therefore, it is better to follow the alternative recipe to obtain a
bounce, a turnaround and in general cyclic behavior,  namely to acquire
extra terms in the gravitational side of the Friedmann equations
for $H^2$ and $\dot{H}$,  in order for the aforementioned
requirements to be fulfilled. Thus,  we need to construct
suitable generalizations of General Relativity,  that are capable
of inducing such novel terms. In the present work we will
describe two possible scenarios to obtain cyclic behavior.
Particularly, these two models are free of the Phantom Menace and
the perturbations are able to pass through the bouncing point
smoothly and without pathology.

\section{Phantom Menace of Cyclic Cosmology in General
Relativity}
\label{problems}

Before proceeding to our modified-gravity cyclic scenarios,   in
this section we briefly show why the cyclicity realization in
the context of General Relativity is problematic. The simplest
scenario of a cyclic universe in the frame of Einstein gravity
is constructed in terms of the double-field quintom model
(Xiong et al. 2008). The first field is a canonical one,
with a positive kinetic term,  called Quintessence, while the second
possesses a negative kinetic term and it is called
Phantom.\footnote{This is equivalent to the Lee-Wick model in
particle physics,  which involves higher derivative terms and
thus
can yield a bouncing solution  when applied into cosmology
(Cai et al. 2009a).} Its dynamics can be described by an action
of
the form
\begin{eqnarray}\label{action_matter}
 S=\int d^{4}x \sqrt{-g} \left [
\frac{1}{2}\partial_{\mu}\phi\partial^{\mu}\phi
-\frac{1}{2}\partial_{\mu}\psi\partial^{\mu}\psi
-V\left(\phi, \psi\right) \right ],
\end{eqnarray}
where $\phi$ and $\psi$ are respectively the canonical and
phantom fields,  and the flat Friedmann-Robertson-Walker (FRW)
metric reads as $ds^2=dt^2-a^2(t)dx^idx^i$. In the framework of
FRW cosmology we can obtain the energy density and  pressure of
the scenario as
\begin{eqnarray}\label{energypressure}
 \rho=\frac{1}{2}(\dot\phi^2-\dot\psi^2)+V(\phi, \psi), ~
 p=\frac{1}{2}(\dot\phi^2-\dot\psi^2)-V(\phi, \psi).
\end{eqnarray}
These two quantities determine the evolution of the universe
through the Friedmann equations
\begin{eqnarray}\label{FRWeqs}
 H^2=\frac{\rho}{3M_p^2}~, ~
 \dot{H}=-\frac{\rho+p}{2M_p^2}~,
\end{eqnarray}
where $M_p$ is the reduced Planck mass (in the following
we use the convention $M_p = 1/\sqrt{8\pi G}$, $c=1$
and $\hbar=1$). Finally, the two scalars satisfy the Klein-Gordon
equations.

Phenomenologically,  a general form of the potential for a
renormalizable model includes operators with dimension 4 or
less,
consisting of various powers of the scalar fields. We impose a
$Z_2$ symmetry,  that is the
potential remains invariant under the simultaneous
transformations $\phi\rightarrow-\phi$ and
$\psi\rightarrow-\psi$. For a detailed quantitative study we
assume the potential form
\begin{equation}\label{potential}
 V\left(\phi, \psi\right) = \left(\Lambda_0 +\lambda\phi\psi
\right)^{2} +\frac{1}{2}m^{2}\phi^{2}
-\frac{1}{2}m^{2}\psi^{2}~,
\end{equation}
where $\lambda$ is a dimensionless constant describing the
interaction between the two fields,  and $\Lambda_0$ a constant
with dimension of $[{\rm mass}]^2$.

In our scenario the two scalars dominate the universe
alternately. In order to present this behavior more
transparently,  we assume,  without loss of generality,  that the
universe starts an expansion from a bounce point. It is then
phantom-dominated and its energy density increases. However,  a
quintessence-dominated stage follows,  and the energy density reaches a
maximal value, after which it decreases. When it arrives at zero,  the
turnaround is realized and the universe enters into a contracting phase.
Therefore,  from the bounce to the turnaround the universe needs
to transit from the phantom-dominated phase into the
quintessence-dominated one, or vice versa. In order to
describe the whole evolution explicitly,  in Fig. 1 we
provide the evolutions for the energy density and the scale
factor of such a cyclic universe. We observe that during each
cycle the universe is dominated by Quintessence and Phantom
alternately.
\begin{figure}[htp]
\centering
\includegraphics[scale=1.15]{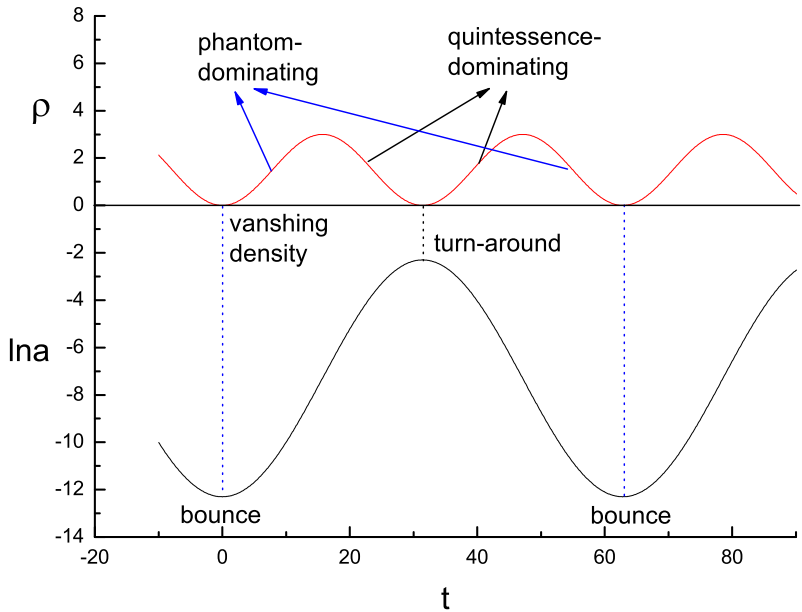}
\caption{{\it{The evolution of a (symmetric) cyclic universe
(Xiong et al. 2008). The scale factor of the universe oscillates
between the minimal and maximal value. For each cycle the
quintessence-like and phantom-like components dominate
alternately.
}}}\label{draft}
\end{figure}

However,  this model suffers from a severe problem of quantum
instability due to the Phantom field
(Carroll et al. 2003, Cline et al. 2004). It is well known
that a
quantum field theory can be well defined based on the existence
of a stable vacuum. In such a vacuum,  pairs of virtual particles
can be produced through quantum fluctuations and then annihilate
rapidly. This implies that,  for all fields,  the kinetic terms
ought to be positively defined,  so that the virtual particles
will not become real ones and the stability of the vacuum will be
preserved. Thus,  once a Phantom field with a negative kinetic
term is introduced,  the vacuum becomes unstable. Even assuming
that ghosts interact gravitationally,  the vacuum is able to
produce a pair of virtual gravitons and then decay into a pair of
ghosts and a pair of other particles such as photons. In this
case,  one can observe that gravitational radiation can be
emitted
from ``nothing" and its density grows exponentially. In summary,
it seems that cyclicity in the context of Einstein gravity, and
thus under the necessary introduction of Phantom fields, suffers
inherently from ghost instabilities\footnote{We mention here that in the
context of General Relativity one may alternatively remove the Big Bang
singularity and enable a cyclic cosmology by introducing a nonlinear
electro-dynamical Lagrangian (De Lorenci et al. 2002) in the
right-hand-side of field equations. This scenario was analyzed in
astrophysics in detail (see e.g. (Mosquera Cuesta \& Salim 2004a, Mosquera
Cuesta \& Salim 2004b)), and its application to black hole physics was
studied in (Corda \& Mosquera Cuesta 2010). }. The plausible way out is to
extend to gravitational theories beyond General Relativity.

\section{Cyclicity in a model of non-relativistic quantum
gravity}
\label{model}

Motivated by a recent work  (Horava 2009),  one realizes
that a scenario of power-counting renormalizable gravity may be
achieved just by adding higher-order spatial derivative terms. The original
model, which is the so-called Ho\v{r}ava-Lifshitz gravity,
suffers from problems such as the over-constraining in UV
region, being not compatible with current observations even in the IR
limit. Additionally, painstaking latest analysis of cosmic Gamma Ray Bursts
impose very stringent constraints on the Lorentz violation  (Laurent et al.
2011), which is a key element of  Ho\v{r}ava-Lifshitz gravity.

The logic of
an effective field theory suggests that a complete action of gravity could
include all possible terms consistent with the imposed symmetries, and the
dimensions of these terms ought to be bounded due to renormalization. In
the frame of 4-dimensional spacetime,  a renormalizable term may
allow for 6th-order spatial derivatives at most (Horava 2009). As
a sacrifice,  the Lorentz symmetry has to be abandoned,  but it
may appear as an emergent one at low-energy scales.

From these we deduce that one could use a modified action which
involves all the permitted terms
(Cai \& Saridakis 2009, Saridakis 2010):
\begin{equation}\label{SgNR}
 \Delta{S}_g = \frac{1}{16\pi G} \int dt d^3x \sqrt{g}N \left(
\alpha_1\bar{R}_{ij}\bar{R}^{ij}
+\alpha_2\bar{R}^2+\alpha_3\nabla_i\bar{R}_{jk}\nabla^i\bar{R}^{jk}
+\alpha_4\nabla_i\bar{R}_{jk}\nabla^jR^{ki}+\alpha_5\nabla_i\bar{R}
\nabla^i\bar{R} \right),\
\end{equation}
where $K_{ij}$ is the extrinsic curvature and $\bar{R}$ is the
three-dimensional Ricci scalar. The dynamical variables are the
lapse and shift functions,  $N$ and $N_i$ respectively,  and the
spatial metric $g_{ij}$ (roman letters indicate spatial indices)
writes as $ds^2= -N^2 dt^2 +g_{ij}(dx^i + N^i dt )(dx^j + N^j dt
)$. Focusing on cosmological frameworks with an FRW metric we use
$N=1$,  $g_{ij}=a^2(t)\gamma_{ij}$,  $N^i=0$,  with
$\gamma_{ij}dx^idx^j = \frac{dr^2}{1-k{r}^2} +r^2d\Omega_2^2 $,
where $k=-1, ~0, ~1$ corresponds to open,  flat,  and closed
geometry
respectively. Finally,  we can insert a matter component in the
scenario,  namely a canonical scalar $\phi$ with action
\begin{eqnarray}
\label{Sm}
 S_m=\int dt d^3x \sqrt{g}N
[\frac{1}{2}\partial_\mu\phi\partial^\mu\phi-V(\phi)]
\end{eqnarray}
into action (\ref{SgNR}),  with $V(\phi)$ its potential,
$\rho_m\equiv\frac{\dot\phi^2}{2}+V(\phi)$ its energy density and
$p_m\equiv\frac{\dot\phi^2}{2}-V(\phi)$ its pressure. We stress
that the above construction is free of Phantom fields.

By varying $N$ and $g_{ij}$ we obtain the Friedmann equations
\begin{eqnarray}
\label{Fr1}
 H^2&=&\frac{8\pi G}{3}\Big[\rho_m+\rho_{k}+\rho_{dr}\Big], \\
\label{Fr2}
 \dot{H}+\frac{3}{2}H^2&=&-4\pi
G\left[p_m-\frac{1}{3}\rho_{k}+\frac{1}{3}\rho_{dr}\right],
\end{eqnarray}
 which are effectively the same as (\ref{FRWeqs}),  but with the
extra contributions of the curvature term
$\rho_k\equiv-\frac{3k}{8\pi Ga^2}$,  and of the term:
\begin{eqnarray}\label{rhodr}
 \rho_{dr}&\equiv&-\frac{3(\alpha_1+3\alpha_2)k^2}{4\pi Ga^4}~.
\end{eqnarray}
This term is the so called ``dark radiation",  since it possesses
effective equation-of-state parameter  $w=1/3$,  that is the same
as normal radiation,  but it exhibits a negative energy density
when $\alpha_1+3\alpha_2>0$. Note that it exists only when the
spatial geometry is not flat (Brandenberger 2009b). In
summary,  in the scenario at hand we obtain a component of
negative energy density,  and thus able to drive cyclic
behavior,
of pure gravitational origin,  without the use of any Phantom
field and therefore free of the ghost instability problem.

Observing the Friedmann equations,  and having in mind the
general
requirements for the realization of cyclic behavior described in
the Introduction,  it is easy to see that these can be easily
fulfilled in the model at hand. In particular,  we have two
choices. The first is to consider a specific matter potential
$V(\phi)$,  and then try to determine the parameter space that
allows for the fulfillment of the requirements. However,  it
would
be better to suitably determine (reconstruct) $V(\phi)$ in order
to acquire cyclicity independently of the parameter values.
Therefore,  imposing a known scale factor $a(t)$ possessing an
oscillatory behavior,  both $H(t)$ and $\dot{H}(t)$ are
straightforwardly known,  and thus we can use the Friedmann
equations together with the pressure and energy density of the
scalar field,  in order to extract the relations for $\phi(t)$.
Finally,  eliminating time between $\phi$ and $V$ we extract the
explicit profile of the potential $V(\phi)$.  $V(\phi)$ will have
an oscillating form too,  since a non-oscillatory $V(\phi)$ would
be physically impossible to generate an infinitely oscillating
scale factor and a universe with a time-symmetry.

The aforementioned bottom to top approach was enlightening about
the form of the scalar potential that leads to a cyclic
cosmological behavior. Alternatively,  we can perform the above
procedure the other way around,  starting from a specific
oscillatory $V(\phi)$ and resulting to an oscillatory $a(t)$. As
a specific example we consider the simple case
\begin{equation} \label{Vtspec}
 V(\phi)=V_0\sin(\omega_V\,  \phi)+V_c.
\end{equation}
In this case,  the Friedmann equation (\ref{Fr2}) gives a
differential equation for the scale factor which can be easily
solved numerically. In Fig. 2 we depict the corresponding
solution for $a(t)$ (and thus for $H(t)$) with $V_0=5.25$,
$\omega_V=0.25$ and $V_c=5.25$,  with $k=1$,  $\alpha_1=1$ and
$\alpha_2=1$ (in units of $M_p$). The potential parameters have
been chosen in order to acquire a cyclic universe with
$a\approx1$ at the bounce.

\begin{figure}[ht]
\begin{center}
\mbox{\epsfig{figure=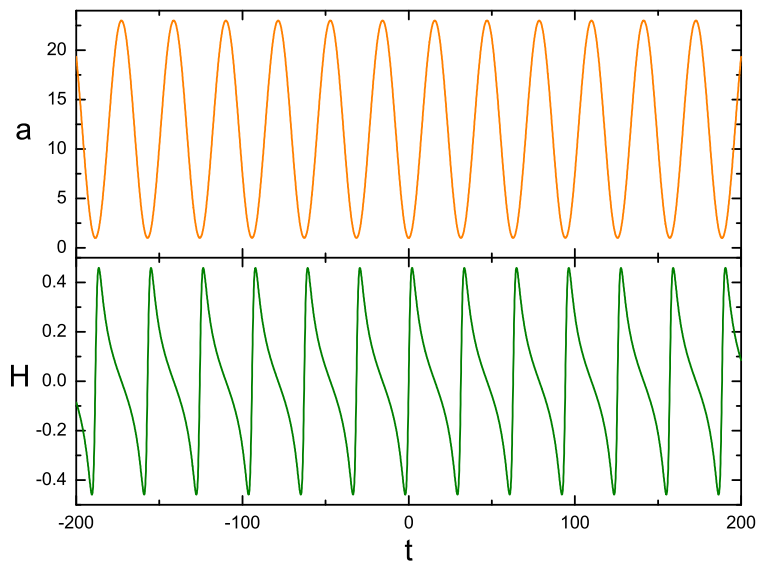,  width=11.5cm,
angle=0}}
\caption{{\it
The evolution of the scale factor $a(t)$ and of the Hubble
parameter $H(t)$,  for a scalar potential of the ansatz
(\ref{Vtspec}) with $V_0=5.25$,  $\omega_V=0.25$ and $V_c=5.25$,
with $k=1$,  $\alpha_1=1$ and $\alpha_2=1$
(Cai \& Saridakis 2009, Saridakis 2010). }}
\label{nr}
\end{center}
\end{figure}
Note that the above specific example is just a simple
representative of cyclic behavior in our gravitational and
cosmological construction,  and it corresponds only to a
sub-class
of the whole set of cyclic models. Obviously,  one can
straightforwardly generalize the aforementioned procedure in any
periodic model.
Moreover,  since the bounce solutions arise owing to the presence
of a dark radiation component with negative energy density,  they
can also be obtained if ordinary radiation with positive energy
density is present. When ordinary radiation is involved,  it has
to be generated from the reheating process of a primordial field
such as the inflaton,  or straightaway from the matter field
$\phi$. Therefore,  its domination  takes place only after
reheating,  of which the energy scale is much lower than the
bounce scale,  and thus the bounce remains unaffected. Finally,
in the late-time evolution, normal radiation would be erased during
matter-dominated period,  and hence it will not affect the bounce
solution in the next cycle.

We close this section by stressing that the scenario at hand is
free of the ghost problem,  since the higher-order derivatives
are only introduced along spatial directions but not in the time-like ones.
Furthermore,  it does not suffer from the strong coupling problem
which exists in Ho\v{r}ava-Lifshitz gravity,  since the kinetic
term for gravity is exactly the same as that of Einstein-Hilbert
action. As a consequence,  we conclude that there are no extra
degrees of freedom. These features are significant advantages,
necessary for a consistent phenomenological description of the
overall universe.

\section{Cyclicity in modified gravity with Lagrange multipliers}
\label{modgrav}

Let us now present cyclicity  in a different modified gravity
scenario,  namely under Lagrange multipliers (Mukhanov \& Brandenberger
1992, Lim et al. 2010,  Capozziello et al. 2010). For simplicity we
focus on the flat FRW geometry (Cai \& Saridakis 2011),  although we could
straightforwardly generalize our results to the non-flat case
too.

We start with a conventional $f(R)$-gravity,  and we add a scalar
field $\lambda$ which is a Lagrange multiplier. In particular,
the action reads:
\begin{equation}
\label{Lag1fr}
 S = \int d^4 x \sqrt{-g} \left\{ f_1(R) - \lambda \left[
\frac{1}{2} \partial_\mu R \partial^\mu R + f_2 (R)
\right]\right\},
\end{equation}
where $f_1(R)$ and $f_2(R)$ are two independent functions of the
Ricci scalar $R$. Note that in the above action we have not
included the matter content of the universe for simplicity,
since
this would significantly modify the multiplying terms of
$\lambda$,  making the subsequent reconstruction procedure
technically very complicated.

Due to the  Lagrange multiplier form,  variation over $\lambda$
leads to an important constraint,  namely
$f_2(R)=\frac{1}{2}{\dot R}^2$.
For $f_2(R)>0$,  this constraint can be solved as
$t = \int^R
\frac{dR}{\sqrt{2f_2(R)}}$
and inverting this relation with respect to $R$ one can obtain
the
explicit $R(t)$. Thus,  using also the definition of the Ricci
scalar one obtains a
differential equation in terms of $H(t)$,  namely
\begin{equation}
 \label{diffeqfr}
 6 \dot{H}(t) + 12 [H(t)]^2 =R(t),
\end{equation}
the solution of which determines completely the cosmological
behavior.

In order to provide a simple realization of cyclicity in this
scenario,  we start by imposing a desirable form of $H(t)$ that
corresponds to a cyclic behavior. In particular,  having
described
the general requirements for cyclicity in the Introduction,  we
choose  $H(t)$ to be straightaway a sinusoidal function:
$H(t)=A_H\sin(\omega_H t)$,  which gives rise to a non-singular
and oscillating scale factor of the form of
\begin{equation} \label{atfr}
 a(t)=A_{H0}\, \exp\left[-\frac{A_H\cos(\omega_H
t)}{\omega_H}\right].
\end{equation}
Thus,  inserting this relation into (\ref{diffeqfr})  we can
easily reconstruct the form of $f_2(R)$ as
\begin{eqnarray}
\label{f2rreconexample}
 f_2(R)=
 \frac{1}{4}\omega_H^2\Big(48A_H^2-4R+3\omega_H^2\Big)
\left[2R-3\omega_H^2-\omega_H\sqrt{3(48A_H^2-4R+3\omega_H^2)}\right].\
\end{eqnarray}

In summary,  such an ansatz for $f_2(R)$ produces the cyclic
universe with scale factor (\ref{atfr}). Note that $f_2(R)$ has a
remarkably simple form,  and this is an advantage of the scenario
at hand,  since in conventional $f(R)$-gravity one needs very
refined and complicated forms of $f(R)$ in order to reconstruct a
given cosmological evolution. However,  the absence of matter
evolution in a cyclic scenario is a disadvantage,  since we
cannot reproduce the epoch evolution of the universe. Therefore it would
be necessary to extend the above formalism under matter
inclusion, similarly to the case of non-relativistic cyclic
cosmology,  a procedure which proves quite complicated
(Cai \& Saridakis 2011). We mention here that a cyclic model of modified
gravity involving Lagrange Multiplier was also studied in the context of
nonlinear electro-dynamical system in (Corda 2008), and its inflationary
solution was addressed in (Corda \& Mosquera Cuesta 2011) (see also
Starobinsky 1979).

We close this section referring to the absence of quantum
instabilities in the scenario at hand. In principle,  one may
worry since the action (\ref{Lag1fr}) involves higher derivative
terms,  which could imply such instabilities. However,  this is
not the case since these annoying terms can be frozen by the
perturbed constraint equation. Additionally,  the vanishing of
the $(i, j)$ component of the perturbed Einstein equation allows to
eliminate one degree of freedom,  and therefore there is still
only one mode of metric perturbation which is able to propagate
freely. In particular,  we assume that a bouncing phase is
realized slowly,  and then the universe approaches a static phase
around the bounce asymptotically. Under this assumption we find
that the kinetic term of the perturbation is positively defined
and canonical,  and that the sound speed of the perturbation is
unity,  as it can be read from the coefficient before the
gradient term  (Cai \& Saridakis 2011). Furthermore,  in such a case
we explicitly confirm that there exist only a single degree of
freedom in Lagrange-multiplier modified $f(R)$ cosmology. In
conclusion,  the scenario at hand is free of instabilities,  and
thus it could be a candidate for the description of the overall universe.

\section{Fluctuations through the bounce}
\label{Fluctuations}

A scenario of non-relativistic gravity is usually able to recover
Einstein's General Relativity as an emergent theory at low-energy
scales. Therefore,  the cosmological fluctuations generated in
this model should be consistent with those obtained in standard
perturbation theory in the IR limit (Brandenberger 2009b).
This result has been intensively discussed in the literature (see
e.g.  Cai et al. 2009a). In particular,  the perturbation
spectrum presents a scale-invariant profile if the universe has
undergone a matter-dominated contracting phase
(Cai \& Zhang 2009a,  Cai et al. 2009b,  Cai et al. 2009c, Cai
et al. 2011). However,  the
corrections in the modified action of the present work could lead
to a modification of the dispersion relations of perturbations.
This issue has been addressed in  (Cai \& Zhang 2009b),
which shows
that the spectrum in the UV regime may have a red tilt in a
bouncing universe. Moreover,  the perturbation modes cannot even
enter the UV regime in the scenario of matter-bounce. Thus,  the
analysis of the cosmological perturbations in the IR regime is
quite reliable. Finally, we mention that the gravitational wave
astronomy may also give a hope to the potential test of modified gravity
models (see Corda 2009).

Things become complicated but more interesting in a cyclic
scenario. Usually,  a particular perturbation mode in the
contracting phase is dominated by its growing tendency,  but in
the expanding stage it becomes nearly constant on super-Hubble
scales. Therefore,  the metric perturbation is amplified on
super-Hubble scales cycle by cycle (Piao 2009, Zhang et al. 2010),  and
moreover the slope of its spectral index is varying (Brandenberger 2009a).
However, it is known that the contribution of fluctuations has to be much
less than the background energy. This prohibits the metric
perturbations to enter the next cycle if $\delta\rho/\rho\sim
O(1)$,  unless the universe can be separated into many parts
independent of one another,  each of which corresponding to a new
universe and evolving up to next cycle,  then separate again and
so on. In this case,  the model of cyclic universe may be viewed
as a realization of the multiverse scenario
(Erickson et al. 2007,  Lehners \& Steinhardt 2009,
 Piao 2009, Zhang et al. 2010).

\section{Conclusions}
\label{conclusions}

In this article, we have reviewed generic features of
non-singular cyclic cosmology. There exist two possible
approaches to realize cyclicity. The first is to introduce
non-conventional matter fields,  such as quintom,  however such a
scenario typically suffers from the problem of quantum
instability. In order to avoid this undesirable pathology,  a
much more promising recipe along the direction of extended Einstein
gravity is expected. Thus, we constructed two explicit examples
of realizing cyclic behavior without the Phantom menace. However, we
should mention here that the whole discussion lies on the grounds of the
cosmological principle, that is on the assumption that the matter
distribution of the cosmos is homogeneous, which may not be the case
according to latest analysis of Sloan Digital Sky Survey (Thomas et al.
2011). This point deserves further investigation.

The first scenario is based on the introduction of higher-order
derivatives of curvature terms in spatial coordinates,  but
preserving the kinetic term of Einstein-Hilbert action unchanged.
This model can yield an oscillatory scale factor of the universe
without the ghost-presence and strong-coupling problems.

The second scenario is the $f(R)$-gravity including a Lagrange
multiplier field. Under the assumption of slow bounce,  we find
only one canonical degree of freedom of which the sound speed is
unity. Thus, it is possible for the cosmological perturbations to
evolve through the bounces without quantum instability,  too.

In conclusion,  the analysis of the present work indicates that
extending to gravitational theories beyond General Relativity
provides a simple and consistent way for the realization of
cyclic cosmology. However, we must mention here that observationally,
General Relativity is still the most compatible theory, both at the
cosmological and Solar system scales, and that any distinguishable
feature of modified gravity is very difficult to be verified. The
motivation of the above investigation arises from the conceptual level, and
it is just the singularity avoidance, which is a disadvantage of General
Relativity.

\begin{acknowledgments}
Y.F.C. is supported by funds from the Department of Physics at
Arizona State University. Y.F.C thanks McGill University for the
hospitality when this work was initiated.
\end{acknowledgments}

\begin{center}{\Large{
\textbf{References}}}
\end{center}

Biswas,  T.,  Mazumdar,   A.,  Siegel,   W. (2006).
Bouncing universes in string-inspired gravity.

$\ \ \ \ \ \ \ \ \ \ \  $
 JCAP {\bf 0603}, 009.

 Bojowald,   M. (2001).
  Absence of singularity in loop quantum cosmology.
  Phys.\ Rev.\ Lett.\  {\bf 86},

$\ \ \ \ \ \ \ \ \ \ \  $
5227.

Brandenberger,  R.~H.  (2009).
Processing of Cosmological Perturbations in a Cyclic
Cosmology.

$\ \ \ \ \ \ \ \ \ \ \  $
  Phys.\ Rev.\  D {\bf 80},  023535.

Brandenberger,    R.  (2009).
Matter Bounce in Horava-Lifshitz Cosmology.
  Phys.\ Rev.\  D {\bf 80},

$\ \ \ \ \ \ \ \ \ \ \  $
043516.

Brustein,    R.,   Madden,   R. (1998).
A model of graceful exit in string cosmology.
  Phys.\ Rev.\  D

$\ \ \ \ \ \ \ \ \ \ \  $ {\bf 57},  712.

Cai,    Y.~F.,     Saridakis,  E.~N.   (2009).
  Non-singular cosmology in a model of non-relativistic

$\ \ \ \ \ \ \ \ \ \ \  $ gravity.
  JCAP {\bf 0910},  020.

Cai,    Y.~F.,   Saridakis,   E.~N.  (2011).
Cyclic cosmology from Lagrange-multiplier modified
gravity.

$\ \ \ \ \ \ \ \ \ \ \  $
  Class.\ Quant.\ Grav.\  {\bf 28},  035010.

Cai,    Y.~F.,  Zhang,   X.  (2009).
Evolution of Metric Perturbations in Quintom Bounce model.

$\ \ \ \ \ \ \ \ \ \ \  $  JCAP {\bf 0906},  003.

Cai,  Y.~F.,  Zhang,   X.  (2009).
Primordial perturbation with a modified dispersion
relation.

$\ \ \ \ \ \ \ \ \ \ \  $
  Phys.\ Rev.\  D {\bf 80},  043520.

Cai,  Y.~F.,  Brandenberger,  R.,  Zhang,   X.  (2011).
The Matter Bounce Curvaton Scenario.
  JCAP

$\ \ \ \ \ \ \ \ \ \ \  $  {\bf 1103},  003.

 Cai,  Y.~F.,  Qiu,  T.~t.,  Brandenberger,  R.,  Zhang,  X.~m.
(2009).
A Nonsingular Cosmology with

$\ \ \ \ \ \ \ \ \ \ \  $ a Scale-Invariant Spectrum of
Cosmological Perturbations from Lee-Wick Theory.

$\ \ \ \ \ \ \ \ \ \ \  $  Phys.\ Rev.\  D {\bf 80},  023511.

 Cai,  Y.~F.,  Saridakis,  E.~N.,  Setare,   M.~R.,  Xia,  J.~Q.
(2010).
Quintom Cosmology: Theoretical

$\ \ \ \ \ \ \ \ \ \ \  $  implications and
observations.
  Phys.\ Rept.\  {\bf 493},  1.

Cai,  Y.~F.,  Xue,  W.,  Brandenberger,  R.,  Zhang,   X.
(2009).
Non-Gaussianity in a Matter

$\ \ \ \ \ \ \ \ \ \ \  $  Bounce.
  JCAP {\bf 0905},  011.

Cai,  Y.~F.,  Xue,  W.,  Brandenberger,  R.,  Zhang,   X.
(2009).
Thermal Fluctuations and Bouncing

$\ \ \ \ \ \ \ \ \ \ \  $ Cosmologies.
  JCAP {\bf 0906},  037.

Cai,  Y.~F., Chen, S.~-H., Dent, J.~B.,   Dutta, S.,  Saridakis, E.~N.
(2011). Matter Bounce

$\ \ \ \ \ \ \ \ \ \ \  \ \ $  Cosmology with the f(T) Gravity.
[arXiv:1104.4349 [astro-ph.CO]] (to appear in

$\ \ \ \ \ \ \ \ \ \ \  \  \ $  Class. Quant. Grav.)

Cai,  Y.~F.,  Qiu,  T.,  Brandenberger,   R.,  Piao,  Y.~S.,
Zhang,  X. (2008).
On Perturbations of

$\ \ \ \ \ \ \ \ \ \ \  $  Quintom Bounce.
  JCAP {\bf 0803},  013.

Cai,  Y.~F.,  Qiu,  T.,  Piao,  Y.~S.,  Li,  M.,  Zhang,  X.
(2007).
Bouncing Universe with Quintom

$\ \ \ \ \ \ \ \ \ \ \  $ Matter.
  JHEP {\bf 0710},  071.

Capozziello,  S.,  Matsumoto,  J.,  Nojiri,  S.,  Odintsov,
S.~D.
(2010).
Dark energy from modified

$\ \ \ \ \ \ \ \ \ \ \  $  gravity with Lagrange
multipliers.
  Phys.\ Lett.\  B {\bf 693},  198.

Carroll,   S.~M.,  Hoffman,   M.,  Trodden,   M.  (2003).
Can the dark energy equation-of-state

$\ \ \ \ \ \ \ \ \ \ \  $  parameter w be less
than -1?.
  Phys.\ Rev.\  D {\bf 68},  023509.

Chen,  S.~-H.,  Dent, J.~B.,  Dutta, S., Saridakis,  E.~N. (2011).
  Cosmological perturbations in

$\ \ \ \ \ \ \ \ \ \ \  $  $f(T)$ gravity.
  Phys.\ Rev.\  {\bf D83}, 023508.

Cline,  J.~M.,  Jeon,  S.,  Moore,  G.~D. (2004).
The phantom menaced: Constraints on low-energy

$\ \ \ \ \ \ \ \ \ \ \  $ effective
ghosts.
  Phys.\ Rev.\  D {\bf 70},  043543.

  Corda, C. (2008).
  An oscillating Universe from the linearized $R^{2}$ theory of gravity.
  
  $\ \ \ \ \ \ \ \ \ \ \  $ 
  Gen.\ Rel.\ Grav.\  {\bf 40}, 2201.

  Corda, C. (2009).
  Interferometric detection of gravitational waves: the definitive test for General 
  
  $\ \ \ \ \ \ \ \ \ \ \  $ Relativity.
  Int.\ J.\ Mod.\ Phys.\  D {\bf 18}, 2275. 

  Corda, C., Mosquera Cuesta, H.~J. (2010).
  Removing black-hole singularities with nonlinear 
  
  $\ \ \ \ \ \ \ \ \ \ \  $ electrodynamics.
  Mod.\ Phys.\ Lett.\  A {\bf 25}, 2423.

  Corda, C., Mosquera Cuesta, H.~J. (2011).
  Inflation from $R^2$ gravity: a new approach using 
  
  $\ \ \ \ \ \ \ \ \ \ \  $ nonlinear electrodynamics.
  Astropart.\ Phys.\  {\bf 34}, 587.

  De Lorenci, V.~A., Klippert, R., Novello, M., M.~Salim, J. (2002).
  Nonlinear electrodynamics  
  
  $\ \ \ \ \ \ \ \ \ \ \  $ and FRW cosmology.
  Phys.\ Rev.\  D {\bf 65}, 063501.

  Erickson, J.~K., Gratton, S., Steinhardt, P.~J., Turok, N. (2007).
  Cosmic Perturbations Through

  $\ \ \ \ \ \ \ \ \ \ \  $  the Cyclic Ages.
  Phys.\ Rev.\  D {\bf 75},  123507.

Feng,  B.,  Wang,  X.~L.,  Zhang,  X.~M. (2005).
Dark Energy Constraints from the Cosmic Age

$\ \ \ \ \ \ \ \ \ \ \  $  and Supernova.
  Phys.\ Lett.\  B {\bf 607},  35.

Gasperini,  M.,   Veneziano,    G. (1993)
Pre - big bang in string cosmology.
  Astropart.\ Phys.\  {\bf 1},

$\ \ \ \ \ \ \ \ \ \ \  $  317.

Horava,  P. (2009).
 Quantum Gravity at a Lifshitz Point.
  Phys.\ Rev.\  D {\bf 79},  084008.

Hoyle, F., Burbidge, G., Narlikar, J. V. (1993).
A quasi-steady state cosmological model with

$\ \ \ \ \ \ \ \ \ \ \  \ $ creation of matter.
Astrophys.J.  {\bf410},  437.

Hoyle, F., Burbidge, G., Narlikar, J. V. (1994).
Astrophysical deductions from the quasi-steady

$\ \ \ \ \ \ \ \ \ \ \  \ $ state cosmology.
 Mon.Not.Roy.Astron.Soc.  {\bf267}, 1007.

Khoury,  J.,  Ovrut,  B.~A.,  Steinhardt,   P.~J.,  Turok,    N.
(2001).
The ekpyrotic universe: Colliding

$\ \ \ \ \ \ \ \ \ \ \  $  branes and the origin of
the hot big bang.   Phys.\ Rev.\  D {\bf 64},  123522.

Khoury,  J.,  Ovrut,  B.~A.,  Seiberg,  N.,  Steinhardt,
P.~J.,
Turok,    N. (2002). From

$\ \ \ \ \ \ \ \ \ \ \  $  big crunch to big bang.
  Phys.\ Rev.\  D {\bf 65},  086007.

 Laurent, P., Gotz,  D.,  Binetruy, P.,  Covino, S., Fernandez-Soto,  A.
(2011).
Constraints on

$\ \ \ \ \ \ \ \ \ \ \  $  Lorentz Invariance Violation using INTEGRAL/IBIS
observations of GRB041219A.

$\ \ \ \ \ \ \ \ \ \ \  $
 Phys.\ Rev.\  D {\bf 83},  121301.

Lehners,  J.~L.,  Steinhardt,  P.~J. (2009).
Dark Energy and the Return of the Phoenix Universe.

$\ \ \ \ \ \ \ \ \ \ \  $  Phys.\ Rev.\  D {\bf 79},  063503.

Lim,    E.~A.,  Sawicki,   I.,   Vikman,   A. (2010).
Dust of Dark Energy.
  JCAP {\bf 1005},  012.

Martin,  J.,  Peter,  P.  (2003).
Parametric amplification of metric fluctuations through a
bouncing

$\ \ \ \ \ \ \ \ \ \ \  $  phase.
  Phys.\ Rev.\  D {\bf 68},  103517.

Mukhanov,  V.~F.,  Brandenberger,  R.~H. (1992). A Nonsingular
universe.   Phys.\ Rev.\ Lett.\  {\bf 68},

$\ \ \ \ \ \ \ \ \ \ \  $  1969.

  Mosquera Cuesta, H.~J., Salim, J.~M. (2004).
  Nonlinear electrodynamics and the surface redshift
  
  $\ \ \ \ \ \ \ \ \ \ \  $  of pulsars.
  Astrophys.\ J.\  {\bf 608}, 925.

  Mosquera Cuesta, H.~J., Salim, J.~M. (2004).
  Nonlinear electrodynamics and the gravitational
  
  $\ \ \ \ \ \ \ \ \ \ \  $ redshift of pulsars.
  Mon.\ Not.\ Roy.\ Astron.\ Soc.\  {\bf 354}, L55.

Novello,    M.,   Bergliaffa,   S.~E.~P. (2008).
Bouncing Cosmologies.
  Phys.\ Rept.\  {\bf 463},  127.

Piao,   Y.~S. (2009).
 Proliferation in Cycle.
  Phys.\ Lett.\  B {\bf 677},  1.

  Saridakis,  E.~N. (2009).
  Cyclic Universes from General Collisionless Braneworld
Models.
  Nucl.\

$\ \ \ \ \ \ \ \ \ \ \  $  Phys.\  B {\bf 808},  224.

  Saridakis,  E.~N. (2010).
  Horava-Lifshitz Dark Energy.
  Eur.\ Phys.\ J.\  C {\bf 67},  229.

  Saridakis,  E.~N.,  Sushkov,   S.~V. (2010).
Quintessence and phantom cosmology with non-minimal

$\ \ \ \ \ \ \ \ \ \ \  $ derivative coupling.
  Phys.\ Rev.\  D {\bf 81},  083510.

  Saridakis,  E.~N.,   Ward,   J. (2009).
Quintessence and phantom dark energy from ghost D-branes.

$\ \ \ \ \ \ \ \ \ \ \  $  Phys.\ Rev.\  D {\bf 80},  083003.

  Saridakis,  E.~N.,  Weller,  J.~M. (2010).
A Quintom scenario with mixed kinetic terms.

$\ \ \ \ \ \ \ \ \ \ \  $  Phys.\ Rev.\  D {\bf 81},  123523.

Shtanov,  Y.,   Sahni,  V. (2003).
Bouncing braneworlds.
  Phys.\ Lett.\  B {\bf 557},  1.

Steinhardt,  P.~J.,  Turok,  N. (2002).
Cosmic evolution in a cyclic universe.
  Phys.\ Rev.\  D {\bf 65},

$\ \ \ \ \ \ \ \ \ \ \  $  126003.

Steinhardt,  P.~J.,  Turok,  N. (2002).
A cyclic model of the universe.
  Science {\bf 296},  1436.

  Starobinsky, A.~A. (1979).
  Relict Gravitation Radiation Spectrum and Initial 
  
  $\ \ \ \ \ \ \ \ \ \ \  $ State of the Universe.
  JETP Lett.\  {\bf 30} (1979) 682.

  Tolman, R.~C. (1931). On the Problem of the Entropy of the
Universe as a Whole.   Phys.\ Rev.\

$\ \ \ \ \ \ \ \ \ \ \  $   {\bf 37},  1639.

 Tolman,  R. C.  (1934).
  Relativity,  Thermodynamics and Cosmology.
  Oxford University Press,

$\ \ \ \ \ \ \ \ \ \ \  $   Oxford,  UK.

Thomas, S.~A.,  Abdalla, F.~B.,  Lahav, O. (2011).
  Excess Clustering on Large Scales in the

$\ \ \ \ \ \ \ \ \ \ \  $
MegaZ DR7 Photometric
Redshift Survey.
  Phys.\ Rev.\ Lett.\  {\bf 106}, 241301.

Veneziano,    G. (1991). Scale Factor Duality For Classical
And Quantum Strings.
  Phys.\ Lett.\  B

$\ \ \ \ \ \ \ \ \ \ \  $  {\bf 265},  287.

Xiong,  H.~H.,  Cai,  Y.~F.,  Qiu,  T.,  Piao,  Y.~S.,  Zhang,
X.
(2008). Oscillating universe with

$\ \ \ \ \ \ \ \ \ \ \  $ quintom matter.
  Phys.\ Lett.\  B {\bf 666},  212.

Zhang,  J.,   Liu,  Z.~G.,  Piao,  Y.~S. (2010).
Amplification of curvature perturbations in cyclic

$\ \ \ \ \ \ \ \ \ \ \  $
cosmology.
  Phys.\ Rev.\  D {\bf 82},  123505.

\end{document}